\documentclass[preprint,showpacs,preprintnumbers,amsmath,amssymb]{revtex4}

\usepackage{graphicx}
\usepackage{dcolumn}
\usepackage{bm}

\begin{document}

%\preprint{AS-ITP-2012-
%\\arXiv:1205.3849 [hep-ph]}

\title{$\theta_{13}$ and the Higgs mass from high scale supersymmetry}

\author{Chun Liu and Zhen-hua Zhao}
\affiliation{
Institute of Theoretical Physics, Chinese Academy of Sciences, \\
and State Key Laboratory of Theoretical Physics,\\
P. O. Box 2735, Beijing 100190, China}
 \email{liuc@itp.ac.cn, zhzhao@itp.ac.cn}

\date{\today}

\begin{abstract}
In the framework in which supersymmetry is used for understanding
fermion masses rather than stabilizing the electroweak scale, we
elaborate on the phenomenological analysis for the neutrino physics.
A relatively large $\sin{\theta_{13}} \simeq 0.13$ is naturally
obtained.  The model further predicts vanishingly small CP violation
in neutrino oscillations.  While the high scale supersymmetry
generically results in a Higgs mass of about $141$ GeV, our model
reduces this mass to 126 GeV via introducing SU(2)$_L$ triplet fields
which make the electroweak vacuum metastable (with a safe lifetime)
and also contribute to neutrino masses.
\end{abstract}



\maketitle

\section{Introduction}

Nowadays the particle physics is at a crucial stage. Experiments have
been pushing the energy, up to which the standard model (SM) is still
valid, higher and higher. In such a situation, neutrino experiments
deserve more and more attention.  Recently Daya Bay \cite{daya-bay}
and RENO \cite{reno} experiments have established the fact that
$\theta_{13}$ is relatively large.  Such a result was indicated
earlier in T2K \cite{t2k}, MINOS \cite{minos} and  Double Chooz
\cite{chooz} experiments, as well as in the global fitting
\cite{fitting}.  This relatively large $\theta_{13}$ has important
implications on theories relevant to the flavor puzzle.

On the theory side, the SM still has problems, like how to understand
the Higgs mass?  If the flavor puzzle can be understood?  Lacking of
new physics signals, the electroweak (EW) scale may just have an
anthropic origin \cite{5,split}.  The TeV scale supersymmetry (SUSY)
is losing its motivation.  Inspired by
the simplicity and beauty of SUSY, one of the authors (Liu) proposed
to use SUSY for understanding the flavor puzzle \cite{6,7}.
In the model, a family symmetry is introduced,
then only one generation of fermions
acquires masses after EW symmetry breaking. That is right the third
generation.  It is SUSY that provides necessary features to break the
family symmetry.  For the lepton sector, once sneutrino fields get
different vacuum expectation values (VEVs), the family symmetry is
broken and the muon becomes massive.  The electron obtains its mass
only through loops due to soft SUSY breaking effects where the family
symmetry explicitly breaks.  To get neutrino masses small naturally,
soft SUSY masses should be very large $\sim (10^{11}-10^{13})$ GeV.
The effective theory of this high scale SUSY breaking model at the
TeV scale is just the SM.

In this Letter, we emphasize on that a relatively large
$\theta_{13}$ is the
right result of this high scale SUSY model.  Furthermore, the model
predicts vanishingly small CP violation in neutrino oscillations.  In
Ref. \cite{7}, an order $0.1$ $\sin\theta_{13}$ was predicted roughly,
and CP violation in the lepton sector was not discussed.  We will
elaborate the analysis on the phenomenology of neutrino oscillations
with a better approximation.

More importantly, the high scale SUSY generically predicts the SM Higgs
mass to be $141$ GeV (for a large $\tan\beta$ as in our case) \cite{a},
while recent LHC experiments have ruled out this mass, and discovered
that the Higgs mass is $126$ GeV \cite{b}.
To reduce the Higgs mass from $141$ GeV to about $126$ GeV, we make
use of an observation of Ref. \cite{12}, to modify the model by
introducing SU(2)$_L$ triplets at the high scale.  They can change
the Higgs quartic coupling at the high energy boundary to be even
negative.  They also contribute additionally to neutrino masses.

In the next section, the basics of the original model is reviewed.
Sect. III introduces in SU(2)$_L$ triplets.  Neutrino phenomenology is
analyzed in Sect. IV.  The Higgs mass is discussed in Sect. V.  The
summary and discussions are given in the final section.

\section{Review}

Within the framework of SUSY, we introduce the $Z_3$ cyclic family
symmetry among the SU(2)$_L$ lepton doublets $L_i$ ($i=1,2,3$) for
three generations.  All the other fields are trivial representations
of this $Z_3$.  It is then convenient to discuss
physics in terms of the following redefined fields,
\begin{equation}
\begin{array}{lll}
L_\tau^\prime & = &\displaystyle \frac{1}{\sqrt 3}(\sum L_i)\,,\\[3mm]
L_e           & = &\displaystyle \frac{1}{\sqrt 2}(L_1-L_2)\,,\\[3mm]
L_\mu         & = &\displaystyle \frac{1}{\sqrt 6}(L_1+L_2-2L_3)\,.
\end{array}
\end{equation}
Because $L_\tau'$ is invariant under the $Z_3$, in general, it mixes
with the down-type Higgs $H_2$.  $L_eL_\mu$ is the only bilinear $Z_3$
invariant combination of above fields, which is also an SU(2)$_L$
singlet.  The $Z_3$ and gauge
symmetric superpotential is expressed as follows \cite{7},
\begin{equation}
{\mathcal W} \supset y_\tau H_d L_\tau E_\tau^c
+ L_e L_\mu(\lambda_\tau E_\tau^c + \lambda_\mu E_\mu^c)
+ \bar{\mu} H_u H_d \,,
\end{equation}
where $L_\tau$ and $H_d$ denote the physical $\tau$ lepton doublet and
the physical down-type Higgs, they are superpositions of  $L_\tau'$ and
$H_2$.  $E^c$ stands for charged lepton singlets, $H_u$ the up-type
Higgs, $y_\tau$ and $\lambda$'s the couplings, and $\bar{\mu}$ a mass
parameter.  Consequently, the mass matrix for charged leptons has the
following form when Higgs and sneutrinos get VEVs,
\begin{equation}
\label{m-l}
M^l=\left(\begin{array}{ccc}
       0 & \lambda_\mu v_{l_\mu} &\lambda_\tau v_{l_\mu} \\
       0 & \lambda_\mu v_{l_e}   &\lambda_\tau v_{l_e}   \\
       0 & 0                     &y_\tau v_d
\end{array}
\right).
\end{equation}
It can be seen that the $\tau$ mass is related to the Higgs VEV and
the muon mass to sneutrino VEVs.  The electron mass is zero at this
stage, and will become finite only after that the SUSY breaking effect
is considered \cite{6,7}.

Neutrinos are massive because the lepton number is violated.  Sneutrino
VEVs result in only one nonvanishing neutrino mass in this model,
\begin{equation}
\label{3}
M^\nu_0 =\displaystyle\frac{a^2}{M_{\tilde{Z}}}\left(\begin{array}{ccc}
         v_{l_e} v_{l_e}  &v_{l_e} v_{l_\mu}   &v_{l_e} v_{l_\tau}  \\
         v_{l_\mu}v_{l_e} &v_{l_\mu} v_{l_\mu} &v_{l_\mu} v_{l_\tau} \\
         v_{l_\tau} v_{l_e}&v_{l_\mu} v_{l_\tau}&v_{l_\tau} v_{l_\tau}
\end{array}
\right),
\end{equation}
where $a=\sqrt{(g^2+g'^2)/2}$, $g$ and $g'$ are SM gauge coupling
constants.  A naturally small neutrino mass is obtained by taking the
soft SUSY breaking scale $M_{\tilde{Z}}$ to be ($10^{11}-10^{13}$)
GeV.  Note that trilinear R-parity violating terms in this model
have negligible contribution to neutrino masses due to large
sparticle masses.

The low energy effective theory is just the SM.  Our model is a high
scale SUSY breaking one.  In obtaining correct EW symmetry breaking,
there is a Higgs doublet with its mass-squared finely tuned to be
small (the EW scale) and negative.  So it gets a non-vanishing EW
scale VEV.  In terms of our high scale fields, in general,
the above SM Higgs doublet field corresponds to a mixture of scalar
fields of $H_u$, $H_d$, $L_e$, $L_\mu$ and $L_\tau$ in the case of
R-parity violation.  It is equivalent to say that the scalar fields
of $H_u$, $H_d$, $L_e$, $L_\mu$ and
$L_\tau$ have VEVs.  The relative sizes of the VEVs are determined
by the relative sizes of soft parameters ($B\mu$ terms).

Neutrino oscillation experiments reveal that there are at least two
massive neutrinos.  In order to provide realistic neutrino masses,
a singlet superfield $N$ was introduced in Ref. \cite{7}.  While
the realistic lepton spectrum and mixing pattern can be obtained.
However, the Higgs mass is still required to be 140 GeV which is
in conflict with recent LHC results.

\section{SU(2)$_L$ triplets}

Instead of singlet fields, SU(2)$_L$ triplet superfields $T$ and
$\bar{T}$ are introduced in this work.  $T$ and $\bar{T}$ have
hypercharge $2$ and $-2$, respectively, and their mass is about
$10^{13}$ GeV.

The $Z_3$ invariant superpotential involving $T$ and $\bar{T}$
fields is as follows,
\begin{equation}
\begin{array}{lll}
{\mathcal W} & \supset & \tilde{y}^\nu \sum_i \{L_i H_2\} T
      + \tilde{\lambda}^\nu_1 \{L_i L_i\} T
      + \tilde{\lambda}^\nu_2 \{L_1 L_2 + L_2 L_3 + L_3 L_1\} T \\
  & & + \tilde{\lambda}^\nu_3 \{H_2 H_2\} T
      + \lambda^\nu_4 \{H_u H_u\} \bar{T}
      + M_T T \bar{T}\,,
\end{array}
\end{equation}
where $\tilde{y}^\nu $ and $\tilde{\lambda}^\nu$'s are couplings.  The
braces denote that the two doublets form a SU(2)$_L$ triplet
representation.  The Lagrangian of the corresponding soft SUSY
breaking terms includes scalar masses and trilinear scalar
interactions.

In terms of the redefined fields in Eq.(1), the superpotential has the
following form,
\begin{equation}
\label{6}
\begin{array}{lll}
{\mathcal W} & \supset & y^\nu \{L_\tau H_d\} T
      + \lambda^\nu_1 \{L_e L_e + L_\mu L_\mu\} T
      + \lambda^\nu_2 \{L_\tau L_\tau\} T \\
  & & + \lambda^\nu_3 \{H_d H_d\} T
      + \lambda^\nu_4 \{H_u H_u\} \bar{T}
      + M_T T \bar{T} \,.
\end{array}
\end{equation}
Couplings are denoted without tildes in this flavor basis.  In the
above derivation, we have made use of the following observation,
\begin{equation}
\{L_eL_e+L_\mu L_\mu\} \propto \{L_iL_i-(L_1L_2+L_2L_3+L_3L_1)\}
\end{equation}
and
\begin{equation}
\{L_\tau'L_\tau'\}   =   \{L_iL_i+2(L_1L_2+L_2L_3+L_3L_1)\} \,.
\end{equation}
In this basis, the Lagrangian of soft SUSY breaking terms is written as
\begin{equation}
{\mathcal L}_{\rm soft} \supset
 m_T^2 \tilde{T}^\dag \tilde{T}
+ m_{\bar{T}}^2 \tilde{\bar{T}}^\dag\tilde{\bar{T}}
+ B_T M_T \tilde{T}\tilde{\bar{T}}
+ A \lambda^\nu_4 \{h_u h_u\} \tilde{\bar{T}}
+ \tilde{m}_{ab} \{\tilde{l}_a\tilde{l}_b\} \tilde{T} + {\rm h.c.} \,,
\end{equation}
where $\tilde{l}_a$ denotes both sleptons $\tilde{l}_e$,
$\tilde{l}_\mu$, $\tilde{l}_\tau$ and the Higgs
$\tilde{l}_h$ identical to $h_d$.  The soft masses are also taken to be
typically about ($10^{11}-10^{13}$) GeV.

Analyzing the scalar potential relevant to $T$ and $\bar{T}$ fields, we
see that $\tilde{T}$'s have VEVs $\displaystyle \sim \frac{v_u^2}{M_T}$.
Although being very small, the VEVs induce new terms to the neutrino
mass matrix.  To be more accurate,
\begin{equation}
M^\nu_1 \simeq \displaystyle -\frac{\lambda^\nu_4 v_u^2}
        {M_T}\left(\begin{array}{ccc}
        \lambda^\nu_1  &0  &0   \\
        0 &\lambda^\nu_1&0 \\
         0&0&\lambda^\nu_2
\end{array}
\right).
\end{equation}
This part of neutrino mass generation is the so-called type-II seesaw
mechanism \cite{seesaw}.
For simplicity, we assume that $\lambda^\nu_1$ is negligibly small.
This smallness can be understood by taking
$\tilde{\lambda}^\nu_1 \simeq \tilde{\lambda}^\nu_2$ because
$\lambda^\nu_1 \sim \tilde{\lambda}^\nu_1 - \tilde{\lambda}^\nu_2$ while
$\lambda^\nu_2 \sim \tilde{\lambda}^\nu_1 + \tilde{\lambda}^\nu_2$.

Consequently, the Majorana neutrino mass matrix has the following form,
\begin{equation}
\label{11}
M^\nu= M^\nu_0 + M^\nu_1 =
        \displaystyle\frac{a^2}{M_{\tilde{Z}}}\left(\begin{array}{ccc}
        v_{l_e} v_{l_e}  &v_{l_e} v_{l_\mu}   &v_{l_e} v_{l_\tau}  \\
        v_{l_\mu}v_{l_e} &v_{l_\mu} v_{l_\mu} &v_{l_\mu} v_{l_\tau} \\
        v_{l_\tau} v_{l_e}&v_{l_\mu} v_{l_\tau}&v_{l_\tau} v_{l_\tau}
         +x v_u v_u
\end{array}
\right),
\end{equation}
where
$x=\displaystyle\frac{M_{\tilde{Z}}}{a^2}
\displaystyle \frac{\lambda^\nu_2 \lambda^\nu_4}{M_T}$.
This matrix is of rank 2.
It has the same form as that given in Ref. \cite{7} where a singlet
field was introduced.  With such a neutrino mass matrix, we will give a
detailed analysis for neutrino oscillations,
in particular for $\theta_{13}$ and CP violation.
In addition to generating neutrino masses, triplet fields will play a
key role in reconciling high scale SUSY scenarios with the experimental
result for the Higgs mass.

\section{Neutrino mixing and CP violation}

Now we give a detailed analysis for phenomenological consequences about
the neutrino physics.  Experimental neutrino oscillation parameters are
measured as follows \cite{9, daya-bay},
$\Delta m^2_{21}=(7.59 \pm 0.21)\times 10^{-5} eV^2$,
$\Delta m^2_{32}=(2.43 \pm 0.13)\times 10^{-3} eV^2$ and
$\sin^2(2\theta_{12}) = 0.861^{+0.026}_{-0.022}$,
$\sin^2(2\theta_{23}) > 0.92$,
$\sin^2(2\theta_{13}) = 0.088 \pm 0.008$.

Looking at the neutrino mass matrix $M^\nu$.
We will consider the case that
$x v_u v_u \gg v_{l_e}^2 \sim v_{l_\mu}^2 \sim v_{l_\tau}^2$.  Because
$M^\nu$ has only two nonvanishing eigenvalues, and it has two origins,
such a case is reasonable.  While $v_{l_\tau}$ which is $Z_3$ invariant
could be larger than $v_{l_{e(\mu)}}$, in general, the largeness is a
factor of $3$ which is not considered in the analysis.  This is
different from the case of Ref. \cite{7} where a somewhat unnatural
cancellation between $x v_u v_u$ and $v_{l_\tau}^2$ was required.  As a
result, a normal hierarchical neutrino mass pattern is obtained,
\begin{eqnarray}
\begin{array}{lll}\vspace{0.2cm}
m_1 & = & 0 \,, \\[0.2cm]
m_2 & = & \sqrt{\Delta m_{sol}^2}\approx
      \displaystyle\frac{a^2 v_{l_e} v_{l_e}}{M_{\tilde{Z}}}\,,\\[0.2cm]
m_3 & = & \displaystyle\sqrt{\Delta m_{atm}^2}
\approx \frac{a^2 x v_u v_u}{M_{\tilde{Z}}} \,.
\end{array}
\end{eqnarray}
The matrix that diagonalizes $M^\nu$ has a simple form,
\begin{equation}
\label{3a}
U_\nu\approx\displaystyle\left(\begin{array}{ccc}
   \displaystyle\frac{r}{\sqrt{1+r^2}} & \displaystyle\frac{1}
   {\sqrt{1+r^2}}   &
      \displaystyle\sqrt{\frac{\Delta m_{sol}^2}{\Delta m_{atm}^2}} \\
   \displaystyle\frac{-1}{\sqrt{1+r^2}}&
     \displaystyle\frac{r}{\sqrt{1+r^2}}&
     \displaystyle r\sqrt{\frac{\Delta m_{sol}^2}{\Delta m_{atm}^2}}\\
    0&0&1
\end{array}
\right),
\end{equation}
where $r = \displaystyle \frac{v_{l_\mu}}{v_{l_{e}}}$ which should be
$\mathcal{O}(1)$.  Taking
$v_u\sim 245$ GeV, $v_{l_e}\sim 20$ GeV, $M_{\tilde{Z}}\sim 10^{12}$
GeV, and $x \simeq 0.04$, the experimental results of neutrino mass
squared differences can be recovered.

In the charged lepton mass matrix Eq. (\ref{m-l}), we take that
$v_d \sim v_{l_{e(\mu, \tau)}}$, $y_\tau \sim \lambda_\tau \sim 0.1$
and $\lambda_\mu \sim 10^{-2}$.  The eigenvalues are,
\begin{eqnarray}
\begin{array}{lll}\vspace{0.2cm}
m_e   &=      &0 \,,\\[0.2cm]
m_\mu &\approx&\lambda_\mu v_{l_e}\sqrt{1+r^2}
        \displaystyle\frac{y_\tau v_d}
        {\sqrt{y_\tau^2 v_d^2+\lambda_\tau^2v_{l_e}^2(1+r^2)}}\,,
    \\[0.2cm]
m_\tau&\approx&
        \sqrt{y_\tau^2 v_d^2+\lambda_\tau^2v_{l_e}^2(1+r^2)}
    \,.
\end{array}
\end{eqnarray}
The matrix $U_l^\dag$ which diagonalizes the mass squared
matrix $M^lM^{l\dag}$, has the following form,
\begin{equation}
U_l^\dag \approx\left(\begin{array}{ccc}\vspace{0.15cm}
\displaystyle\frac{1}{\sqrt{1+r^2}}&
 \displaystyle\frac{-r}{\sqrt{1+r^2}}&0\\\vspace{0.15cm}
\cdots&\cdots&\cdots\\\vspace{0.15cm}
\cdots&\cdots&\cdots
\end{array}
\right)\,
\end{equation}
where we have just listed the matrix elements which are relevant
for our purpose.

Thus, the lepton mixing matrix $U^{PMNS}\equiv U_l^\dag U_\nu$
\cite{pmns} is obtained.  It is parameterized in the following
standard form,
\begin{equation}
U^{PMNS}=\left(\begin{array}{ccc}
c_{12}c_{13}&s_{12}c_{13}&s_{13}e^{-i\delta}\\
\cdots&\cdots&s_{23}c_{13}\\
\cdots&\cdots&c_{23}c_{13}
\end{array}
\right) \times {\rm diag}(e^{i\alpha}, e^{i\beta}, 1) \,,
\end{equation}
where $\delta$ is the Dirac CP violation phase, $\alpha$ and $\beta$
Majorana CP violation phases.  In the case of $m_1=0$, the phase
$\alpha$ is unphysical.  Consequently,
$\tan{\theta_{12}}=\displaystyle \frac{1-r^2}{2r}$ and
$\sin{\theta_{13}}=\displaystyle \frac{1-r^2}{\sqrt{1+r^2}}
\displaystyle \sqrt{\frac{\Delta m_{sol}^2}{\Delta m_{atm}^2}}$.
With the experimental result of $\tan{\theta_{12}}$, $r$ is
determined to be 0.53.  In this case,
$\sin{\theta_{13}}\approx 0.13$ which is very close to the value
of $0.15\pm0.02$ given by Daya Bay experiment \cite{daya-bay}.

Finally, $\theta_{23}$ can be obtained through the following equation,
\begin{equation}
\tan{\theta_{23}}=\sqrt{1+r^2}\frac{\lambda_\tau v_{l_e}}{y_\tau v_d}\,.
\end{equation}
A large $\theta_{23}$ is quite natural but there is no reason in this
model to have $\theta_{23}$ being exactly $\displaystyle\frac{\pi}{4}$.
Nevertheless even global fitting allows $\theta_{23}$ being as low as
$42^\circ$ at the 1 $\sigma$ level.

Now, let us consider CP violation in the lepton sector.  Looking
at mass matrices (\ref{m-l}) and (\ref{11}), because of the family
symmetry they are very special, all the matrix elements of the Dirac
part can be real by redefining relevant fields.
This can be seen in the following way.  Generally sneutrino VEVs are
complex because they are determined by the scalar potential in which the
soft masses are involved.  In the charged
lepton mass matrix, all the phases can be absorbed into fields of
left-handed and right-handed charged leptons.  In the neutrino mass
matrix, if first we do not consider the $x$ term, it is easy to see
that the phases can be rotated out, namely after factorizing the
Majorana phases out, the remaining part of the neutrino mass matrix is
real.  Then once the $x$ term is added, it is noted that we always have
the freedom to adjust its phase to be the same as that of
$v_{l_\tau}v_{l_\tau}$ through a phase rotation of $H_u$ (and $H_d$)
field(s).  Hence the Dirac part of the neutrino mixing matrix is real,
CP violation in neutrino oscillations vanishes in the symmetry limit.
Things become complicated when the nonvanishing electron mass is
counted.
The electron mass is due to the following terms adding to the mass
matrix (\ref{m-l}) \cite{6,7},
\begin{equation}
\delta M^l_{\alpha\beta}\simeq
\frac{\alpha}{\pi}\frac{y_{\tau}\tilde{m}_S v_d}{M_{\tilde{Z}}}
\end{equation}
with $\tilde{m}_S$ being Yukawa soft masses where CP violating phases
are expected to appear.  $\delta M^l_{\alpha\beta}$ itself is a
general $3\times3$ matrix, it is a perturbation to the mass matrix
(\ref{m-l}), thus the Dirac CP violation phase in the lepton sector is
found as $m_e/m_\tau$.  For CP violation to be measured in neutrino
oscillations,
\begin{equation}
\delta \sim \frac{m_e}{m_\tau} \sim 10^{-3} \,, ~~~ {\rm or}
~~~J \sim 10^{-5} \,,
\end{equation}
where $J$ stands for the Jarlskog invariant \cite{jarlskog}.
It is too small to be observed in current neutrino experiments.
Finally, the effective Majorana neutrino mass
$m_{ee}=\sum U^{PMNS 2}_{ei} m_i$ which is to be measured in
neutrinoless double $\beta$ decays, is $\mathcal{O}(10^{-3})$ eV, its
exact value still depends on a Majorana phase which does appear as
we have seen from the above described procedure of phase rotation
\cite{mee}.  It should be mentioned that the discussion of CP
violation is independent on numerical assumptions adopted in the
neutrino mixing analysis.

It is remarkable to compare the mixing matrix of the lepton sector with
that of the quark sector which has been analyzed in detailed in Ref.
\cite{7}.  For the case of quarks, a $Z_3$ symmetry among the three
generation quark doublets is still there.  It is also the soft Yukawa
trilinear interactions that lead to CP violation.  However, the mass
story is a bit different.  The roles of sneutrino VEVs and loop effects
are switched.  Sneutrino VEVs give a mass to the first generation,
whereas loop effects contribute masses to charm and strange quarks.  And
loop effects appear in both up- and down-type quark masses.  Therefore,
it is expected that the lepton mixing and the quark mixing are
qualitatively different.  The final expression of the CKM matrix is
obtained in Eq. (35) of Ref. \cite{7}, in which we can see that small
$V_{ub}$ and $V_{cb}$, and a large Cabbibo angle are natural,
furthermore, a large CP violation phase can be also naturally there.

\section{Higgs mass}

The real aim of introducing triplet fields $T$ and $\bar{T}$ is for the
Higgs mass.  In the high scale SUSY scenario, the Higgs quartic
coupling evolves from a very high SUSY breaking scale
(say $10^{12}$ GeV) with usual boundary condition
$\displaystyle\frac{1}{8}(g^2+{g^\prime}^2)\cos^2{2\beta}$ \cite{a}
down to the EW scale in the same way as that in the SM.  In the case of
a large $\tan{\beta}$, the Higgs mass of ($141 \pm 2$) GeV was
predicted \cite{a}.  Because of sneutrino VEVs, the value will have a
$1\%$ decrease.  Such a mass has been ruled out by recent results of
ATLAS and CMS which show that the Higgs mass is about $126$ GeV
\cite{b}.  To reduce the Higgs mass from $140$ GeV to about $126$ GeV
or so, we make use of an observation of Giudice and Strumia that
triplet fields can change the boundary condition of the Higgs quartic
coupling by a considerable amount in the case of a large $\tan{\beta}$
\cite{12}.

In Eq. (\ref{6}), after integrating out the triplet fields, we get a
new contribution to the boundary condition of the Higgs quartic
coupling.  The interaction of $\lambda^{\nu}_4$ plays the main role,
because its correction to the Higgs quartic coupling is proportional
to $\sin^4{\beta}$ which is significant in the large $\tan{\beta}$
case,
\begin{equation}
\Delta \lambda \simeq \lambda^{\nu 2}_4 \sin^4{\beta}\left[1-
 \displaystyle\frac{M_T^2(m_\beta^2-2 A B_T) + A^2 m_\alpha^2}
 {{m_\alpha}^2{m_\beta}^2-B_T^2 M_T^2}\right] \,
\end{equation}
where
\begin{equation}
{m_\alpha}^2=m_T^2+M_T^2, ~~~ {m_\beta}^2=m_{\bar{T}}^2+M_T^2.
\end{equation}
With the assumption that $M_T^2$ is much larger than $m_{\rm soft}^2$,
the above equation becomes
$\Delta\lambda\sim\displaystyle\frac{\lambda^{\nu 2}_4 \sin^4{\beta}}{M_T^2}[m_T^2-(B_T-A)^2]$.
Obviously, this contribution is comparable to
$\displaystyle\frac{1}{8}(g^2+{g^\prime}^2)\cos^2{2\beta}$ when
$\displaystyle\frac{\lambda^{\nu}_4 m_T}{a M_T} \sim \frac{1}{2}$.  We
note that this is consistent with our numerical choice for $x$ in the last
section, if $\lambda_2^\nu \sim 0.04$.  Moreover, $\Delta \lambda$ is
possible to become negative in some parameter space so that it cancels
$\displaystyle\frac{1}{8}(g^2+{g^\prime}^2)\cos^2{2\beta}$ and even
makes the
boundary condition negative.  With appropriate negative quartic coupling
(about $-0.02$ as shown in \cite{13})
at the high scale, a Higgs mass of about $126$ GeV compatible with the
recent experimental results can be obtained.

A negative Higgs quartic coupling constant at $\sim 10^{12}$ GeV implies
the instability of the SM vacuum.  Note that at even higher energies,
SUSY restores, the Higgs quartic coupling then becomes positive.
Therefore, the true vacuum is at about $10^{12}$ GeV.  This negative
value of the coupling at the high scale remains in the safe region
where the life-time of the SM vacuum due to
quantum tunneling is longer than the age of the Universe \cite{13a,13}.

\section{Summary}

In conclusion, we have discussed the neutrino phenomenology in the
framework which uses SUSY to account for the fermion mass hierarchy
problem.
One natural result of this model is that $\sin{\theta_{13}}$ is about
$0.7 \displaystyle \sqrt{\frac{\Delta m_{sol}^2}{\Delta m_{atm}^2}}
\approx 0.13$, just consistent with experiments.
Furthermore, it has been predicted that the CP violation effect in
neutrino oscillations is vanishingly small.  Besides, the neutrino
masses possess the normal hierarchy;
$\theta_{23}$ is not necessarily $45^\circ$.  These can be checked in
future experiments.

The SUSY breaking scale is necessarily high in order to get neutrino
masses small naturally.
For reducing the high scale SUSY predicted Higgs mass, we have
introduced SU(2)$_L$ triplet fields.  They make the Higgs quartic
coupling negative at the high scale.  Our universe with the EW scale
$\sim 100$ GeV is metastable, and the true vacuum is at $\sim 10^{12}$
GeV.  Because the vacuum energy in our current universe is tiny
positive, it is natural to guess that the true vacuum has a large
negative cosmological constant.

A special form of the neutrino mass matrix Eq. (\ref{11}) has been
adopted in our analysis.  While this is a reasonable assumption in
the case of triplet fields, it can be made accurate by introducing
both singlets and triplets, and the singlets will be only for neutrino
masses, and the role of triplets will be purely for the Higgs mass
reduction.  This will be achieved by raising the triplet mass $M_T$
and the corresponding soft mass $m_{T(\bar{T})}$ two orders of
magnitude with their ratio unchanged.

In fact, in this model, the scale of $10^{12}$ GeV or so is more
fundamental, the effective theory below this scale is the just the
SM, and the EW scale is a kind of accidental via fine tuning.
Conventional WIMP dark matter, therefore, does not exist.  However, we
note that this high scale is close to that of the axion.
The axion might be
the dark matter in this model.  Cosmological aspects of this model are
under our study.

In the near future, if experiments show that the SM is the full theory
at TeV, then leptonic CP violation $\delta$ will be almost the
last parameter to be fixed for the elementary particle physics.  It
would be also almost the last nontrivial physical quantity to verify
various flavor models.  Where the whole flavor puzzle is the most
complicated problem of the SM, SUSY as well as its breaking just
provides a simple and also complicated enough framework to understand
the puzzle.  SUSY should have a use because of its
beauty and powerfulness.

\begin{acknowledgments}
We would like to thank Jia-shu Lu and Hua Shao for helpful discussions.
This work was supported in part by the National Natural Science
Foundation of China under Nos. 11075193 and 10821504, and by the
National Basic Research Program of China under Grant No. 2010CB833000.
\end{acknowledgments}

\end{document}